\documentclass[10pt]{article}

\usepackage{amsmath}
\usepackage{amssymb}

\usepackage{graphicx}

\usepackage{cite}

\usepackage{color} 

\usepackage[labelfont=bf,labelsep=period,justification=raggedright]{caption}

\makeatletter
\renewcommand{\@biblabel}[1]{\quad#1.}
\makeatother

\date{}

\begin{document}

\begin{flushleft}
{\Large
\textbf{Minimal Bounds on Nonlinearity in Auditory Processing}
}
\\
Jacob N. Oppenheim$^{1}$, 
Pavel Isakov$^{1}$, 
Marcelo O. Magnasco$^{1,\ast}$
\\
\bf{1} {Laboratory of Mathematical Physics, Rockefeller University, New York, New York 10065}
\\
$\ast$ E-mail: magnasco@rockefeller.edu
\end{flushleft}

\section*{Abstract}
Time-reversal symmetry breaking is a key feature of nearly all natural sounds, caused by the physics of sound production.  While attention has been paid to the response of the auditory system to ``natural stimuli," very few psychophysical tests have been performed.  We conduct psychophysical measurements of time-frequency acuity for both ``natural" notes (sharp attack, long decay) and time-reversed ones.  Our results demonstrate significantly greater precision, arising from enhanced temporal acuity, for such ``natural" sounds over both their time-reversed versions and theoretically optimal gaussian pulses, without a corresponding decrease in frequency acuity.  These data rule out models of auditory processing that obey a modified ``uncertainty principle" between temporal and frequency acuity and suggest the existence of statistical priors for naturalistic stimuli, in the form of sharp-attack, long-decay notes.  We are additionally able to calculate a minimal theoretical bound on the order of the nonlinearity present in auditory processing.  We find that only matching pursuit, spectral derivatives, and reassigned spectrograms are able to satisfy this criterion.

\section{Introduction}

\begin{figure}[!ht]
\begin{center}
\includegraphics[width=2in]{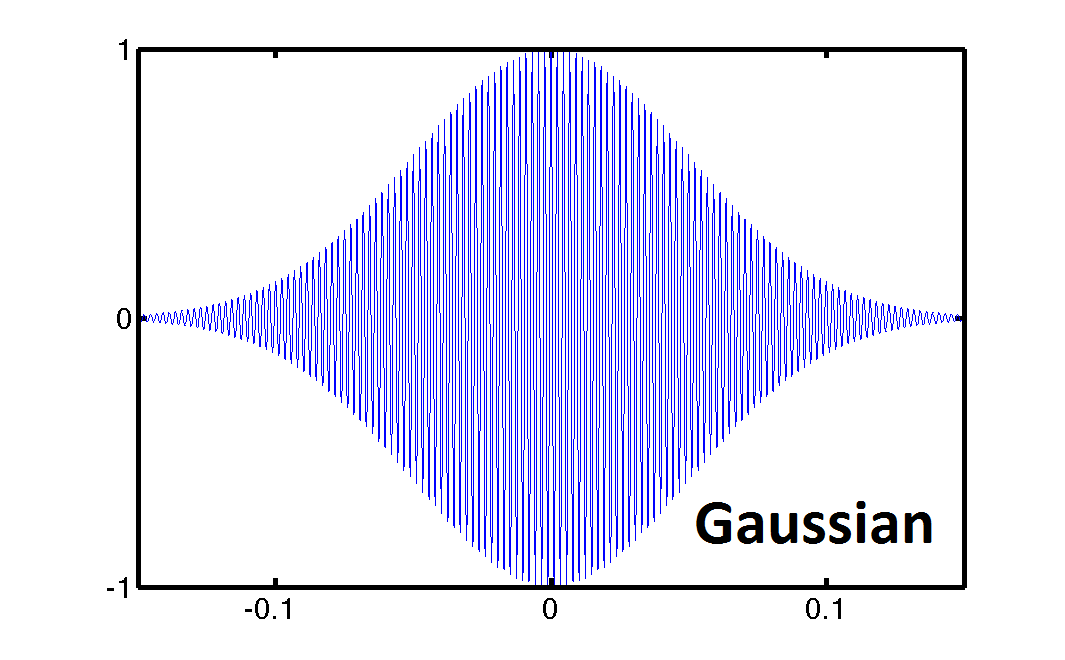}
\includegraphics[width=2in]{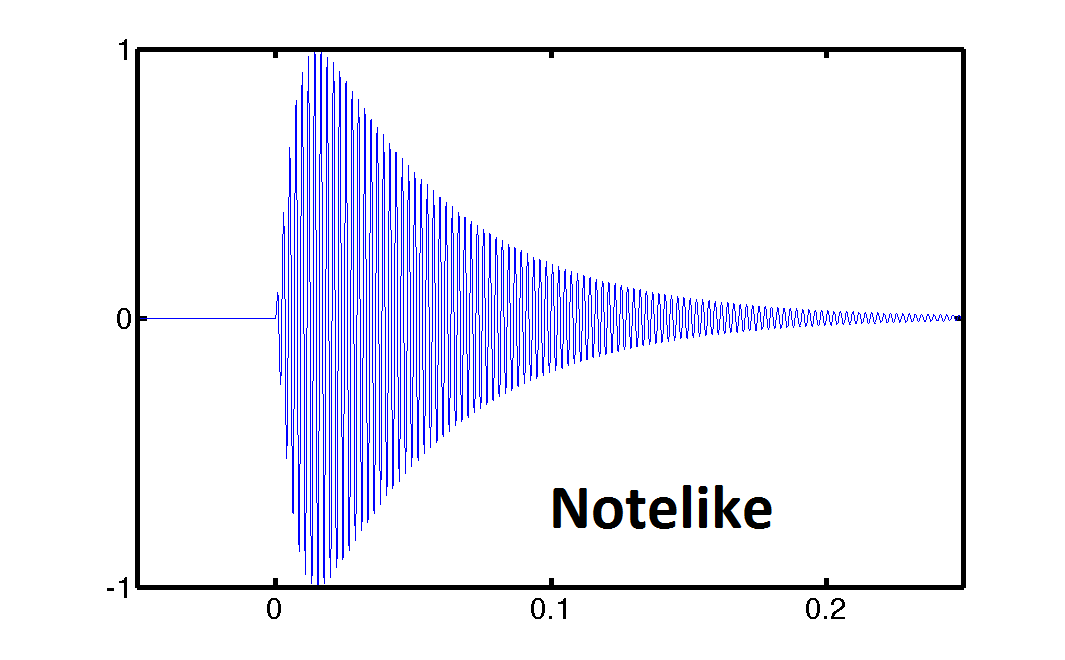}\\
\includegraphics[width=2in]{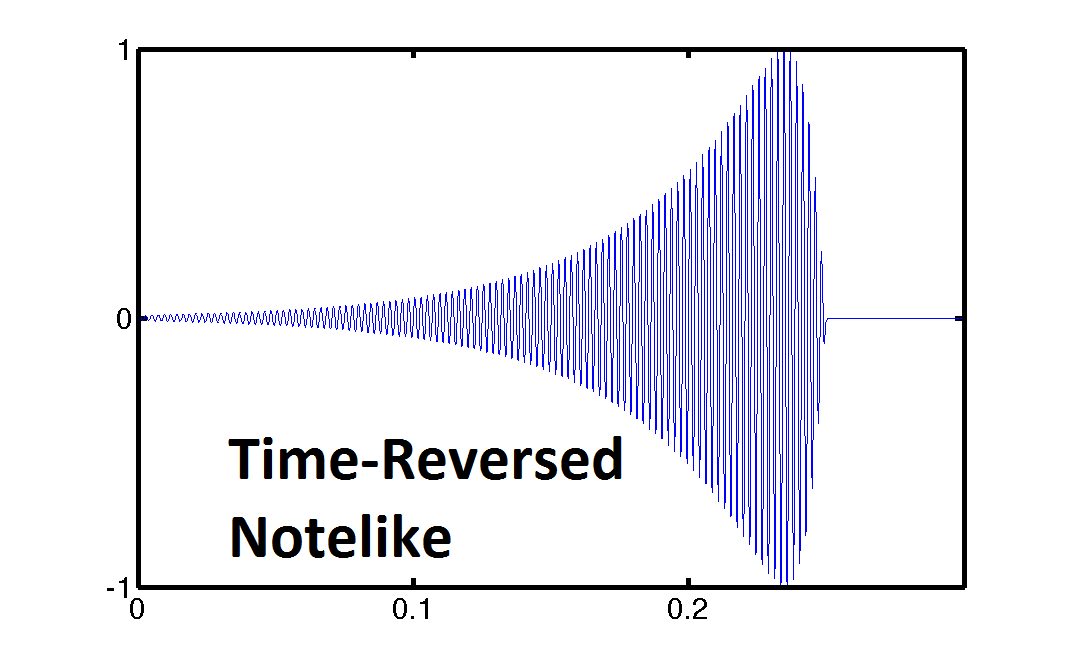}
\includegraphics[width=2in]{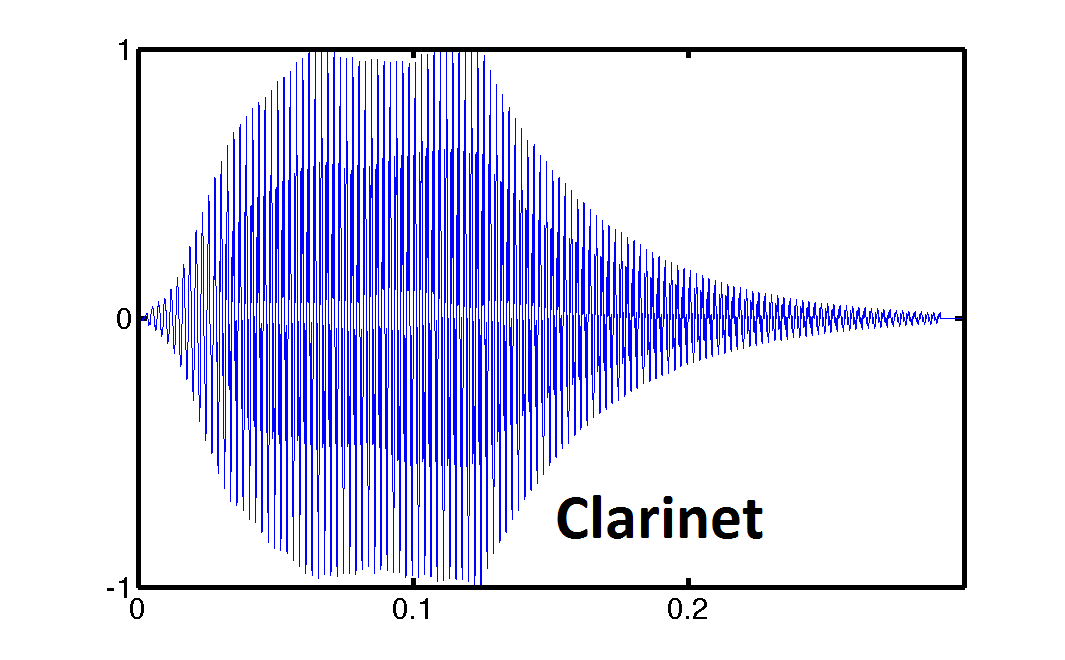}\\
\includegraphics[width=2in]{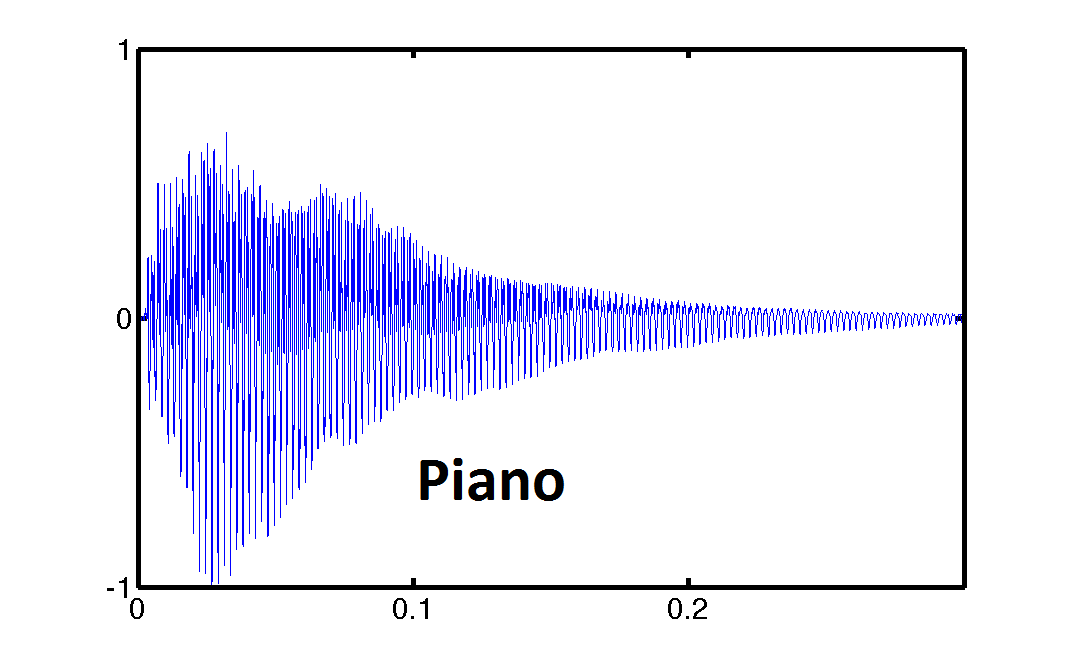}
\includegraphics[width=2in]{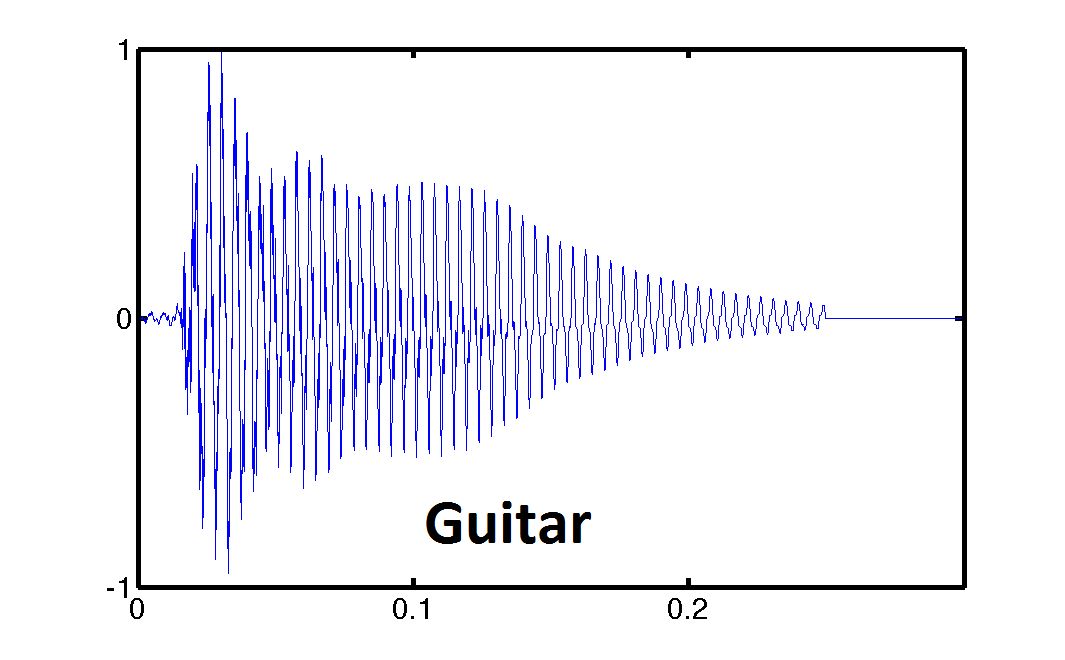} \\
\end{center}
\caption{{\bf Wavepackets used in experiments and musical notes, at 440 Hz.}  Note that each of the musical notes is time-reversal symmetry broken. \label{ExampleNotes}
}\end{figure}

It is natural to question whether the human auditory system is in some way ``optimized" to detect natural sounds.  This principle is clear in the case of the range of frequencies and volumes that humans can hear.  Barlow suggested that the auditory system had evolved to optimally encode natural sounds \cite{Barlow}.  Much recent work in this direction has involved examining the output of neurons in an animal subjected to auditory signals both with natural and unnatural statistics, in amplitude \cite{Rieke, Escabi}, spectrum \cite{Nelken, Woolley, Rodriguez}, and scale-invariance \cite{Geffen}.  While these studies focus on the higher order properties of natural sounds, such as the energy in various spectral bands, comparatively little attention is paid to the lower order properties of such sounds.  Nearly all natural sounds are time-reversal symmetry broken, with sharp attacks and long decays \cite{Geffen}.  This effect may be difficult to detect on spectrograms, as the onset of many notes is so rapid, and the sustain and decay so elongated, that the waveform appears to be a continuous block of sound.  However, from the perspective of parametrizing the possible envelopes of notes, such sounds correspond to a rather small region of parameter space.  One need look no further than the utility and ubiquity of gammatones in scientific research and the ADSR (attack-decay-sustain-release) description of notes in synthesizers.    The second row of Figure \ref{ExampleNotes} demonstrates this property for three different methods of sound production, a clarinet, a piano, and a guitar.

The natural shape of a sharp attack followed by a long decay (or sustain-release) reflects the physics of sound production.  A burst of energy is produced and decays due to natural damping: a mallet hits a drumhead, a burst of air is forced through a trumpet, a string is plucked and released \cite{Geffen}.  Attempts to reconstruct the receptive fields of auditory neurons using the reverse-correlation method have found similar filter shapes \cite{SmithLewicki}.  This is the result we would expect on the grounds of optimal-information transfer: statistical priors of a rapid-attack, slow-decay form.  The reverse-correlation method, however, has only a limited ability to reconstruct auditory filters, even in the case of simulated data \cite{TkacikMagnasco}.  Additionally, such spectral methods are thrown into doubt by the existence of essential nonlinearities in the cochlea \cite{Hopf1, Hopf2} and the recent results that human auditory perception is more precise than any linear (spectral) method can account for \cite{OppenheimMagnasco}.

Unlike other ``higher-order" properties of natural sounds, such as amplitude fluctuations and correlations within the spectrogram, the ability of the auditory system to respond to time-reversal symmetry broken notes versus unnatural ones is easy to test using standard psychophysical methods.  Our previous work had involved the creation of a protocol for measuring simultaneous human time-frequency acuity.  This methodology can easily be adapted to test how both time and frequency acuity change with variations in the envelope of the notes presented.  Such a test would be able to directly infer the existence of time-reversal symmetry broken statistical priors in the auditory system. 

Such statistical priors would help to explain the utility of the numerous nonlinearities found in the auditory system.  Previous work by the authors showed simultaneous time-frequency perception of pure tones was decisively more precise than allowed by the Fourier Uncertainty Principle, by factors of up to 50 \cite{OppenheimMagnasco}.  If we view a signal as a distribution in time and frequency, 
\begin{eqnarray*}
P(t) &=& |x(t)|^2\over{\int_{-\infty}^\infty|x(t')|^2 dt'}			\\
P(f) &=& |\tilde{x}(f)|^2\over \int_{-\infty}^\infty|\tilde{x}(f')|^2 df'	
\end{eqnarray*}
the product of these variances is bounded from below for all linear time-frequency representations:
\begin{eqnarray*}
\Delta t = \sqrt{var(t)} & & \Delta f = \sqrt{var(f)} \\
\Delta t \Delta f &\ge& {1\over 4	\pi }
\end{eqnarray*}
The violation of this bound exposes auditory processing as fundamentally nonlinear: the nonlinearities present in the auditory system allow improved precision.  

When using nonlinear time-frequency distributions it is important to distinguish between two types of uncertainty relation.  The direct analogue of the classical Fourier Uncertainty Principle is one between resolution and precision.  Resolution refers to our ability to distinguish two separate wavepackets, while precision refers to our ability to measure the timing and frequency of a wavepacket we know to be isolated \cite{OppenheimMagnasco}.  Properly defined, the product of resolution and precision is limited from below, yielding the Gabor limit in the classical spectrogram.  We may also investigate the relation between $\Delta t$ and $\Delta f$ for an isolated pulse, providing a quantitative measure of precision. 

Further psychophysics is expected to be able to rule out certain classes of nonlinear models wholesale, in part by taking advantage of the fact that many nonlinear time-frequency representations obey a modified uncertainty principle with a prefactor depending on the order of the nonlinearity \cite{CohenBook}.  Our goal was thus two-fold, to see if the auditory system is primed to better process naturalistic, time-reversal symmetry broken notes, and to use such psychophysical data to better understand the nonlinearities present in the auditory system.

\section{Materials and Methods}
We used the same testing procedure as in \cite{OppenheimMagnasco}.  The relevant specifications of equipment, training tasks, experimental parameters, and preliminary data fitting for the extraction of physiological parameters may be found therein.  What follows is a brief overview of the experimental procedure with emphasis on the adaptations of our original protocol to investigate the effects of time reversal.

{\bf Human subjects. }Our work was a continuation of a prior study \cite{OppenheimMagnasco}, approved under Rockefeller University protocol MAG-0694. As in the previous work, we enrolled highly-musically-trained subjects, due to their superior and stable performance on discrimination tasks, especially those involving simultaneous time-frequency perception.  This study consisted of 12  subjects, many of whom were students of composition or conducting \cite{OppenheimMagnasco}.  Our sample is clearly not representative of the population at large, rather, we desired results directly comparable with previous work involving the same basic auditory task [see below].

{\bf Stimuli and Tasks.}  Three types of wavepackets were used: a Gaussian with a width of 0.05 seconds, a ``notelike'' envelope that approximated a musical note with a rapid increase proportional to $\tanh ^2 (t)$  followed by a slower exponential decay, and a time-reversed version of the second pulse, that is, a gradual attack followed by a sharp decay, see Figure \ref{ExampleNotes}.   The width of the Gaussian was used as the time constant of the exponential in the ``notelike"  and ``time-reversed'' pulses.  We extracted the theoretical uncertainties in time and frequency of each pulse by integration in MATLAB ($\Delta t$ and $\Delta f$); these are the analogous quantities to those extracted from the fitted psychometric curve ($\delta t$ and $\delta f$).    Testing was performed at $f_0=440$ Hz, as this was shown in earlier work \cite{OppenheimMagnasco} to yield optimal performance on the uncertainty task.  The flanking note was separated from the center by a factor $\phi=(\sqrt 5+1)/2$.

{\bf Tasks and testing sequence. } We performed the auditory equivalent of a two-dimensional Vernier task, in which a frame is given specifying a horizontal (time) and a vertical (frequency) direction, and the test note is misaligned from this frame.  Four training tasks were given, including basic frequency and time discrimination tasks.  Subjects performed 5 sets of 20 questions each per task, with additional sets of 20 if necessary for convergence of data, or to eliminate poor initial performance on a task.  All tasks adapted dynamically to the subject's performance according to the two down, one up paradigm.  The difficulty of each task (the amount the test note was misaligned from the flanking notes), $Dt$ and $Df$,  was chosen from a Gaussian distribution. As in our prior work, we adjusted the variance of this distribution, increasing it by $10\%$ after two correct responses and decreasing it by $11\%$ after an incorrect response. Simulations showed these update rules to give the most even sampling of the ``steep" region of the psychometric curve and yield rapid convergence of parameters.  We set up both subtasks as a 2AFC (two alternative forced choice) asking whether the test note comes before or after the high note, and is above or below the test note.  We fit a psychometric function of form erfc(($x$-$a$)/$b$), implicitly assuming that the probability of not noticing a difference was normal in the difference.

\section{Psychophysical Measurements}
Upon testing subjects with all three pulses, it was immediately apparent that performance on the time-reversed task was notably worse.  While several subjects managed relatively similar performance on the preliminary frequency and time discrimination tasks, comparing the notelike (subscript N) and the time-reversed pulses (subscript TR), no one managed better, or even similar performance on the combined time-frequency discrimination task.  We tested the significance of this the ratios of the uncertainty products, $\frac{\delta f_{TR} \delta t_{TR}}{\delta f_{N} \delta t_{N}}$ as well as the limens of discrimination themselves, $\delta f_{TR}/\delta f_{N}$ and $\delta t_{TR}/\delta t_{N}$.  The ratio of the performances was chosen over the difference, both to conform to the Weber-Fechner Law, and because changes in $\delta{f}$ and $\delta{t}$ are bounded from below; one cannot have a negative limen of discrimination.  We used a paired t-test on the difference of the logarithms, i.e. $\log{\delta f_{TR}}-\log{\delta f_{N}}$, implicitly testing for mean 1 lognormality.  For the uncertainty products, we found $P < 0.01$ that the results arose from the same distribution.  That is, the ratio of the performance in uncertainty product between the two tasks indicated a significant fold-change in performance.  

\begin{figure}[!ht]
\begin{center}
\centerline{\includegraphics[width=4in]{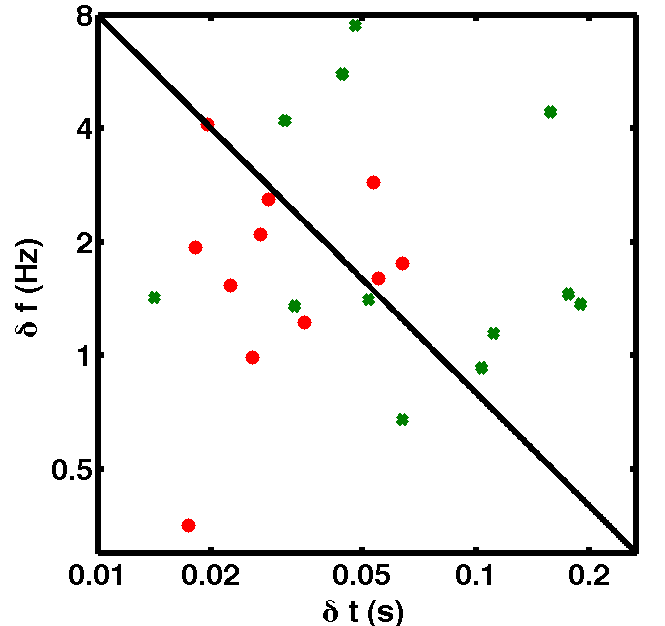}}
\end{center}
\caption{{\bf Results of Psychophysical Testing.}  We plot our data on a standard time-frequency plane with $\delta t$ on the x axis and $\delta f$ on the y axis. The red circles are the results from the ``time-forward" notelike pulse, the green crosses are from the ``time-reversed" version.  The uncertainty bound is the black diagonal; the axes are logarithmic.  The data from the ``time-reversed" pulse are shifted notably to the right ($+\delta t$) from the original ``time-forward" notelike one. \label{TF Plot}
}\end{figure}

\begin{figure}[!ht]
\begin{center}
\includegraphics[width=2in]{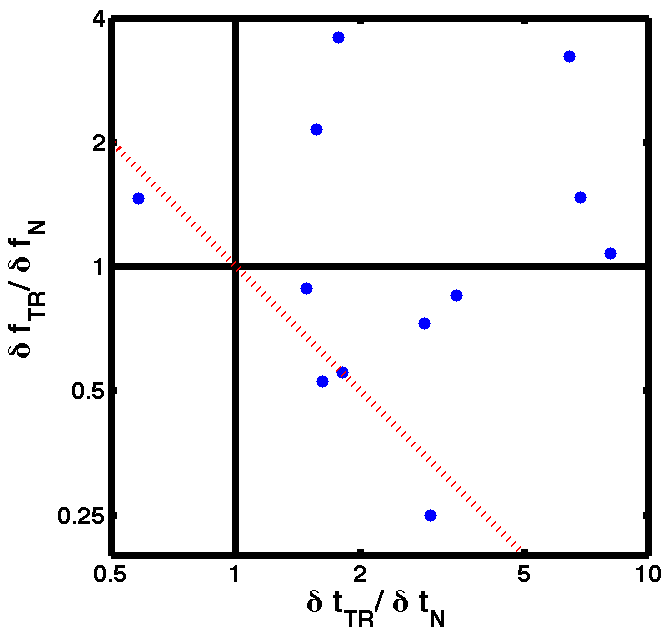}
\includegraphics[width=2in]{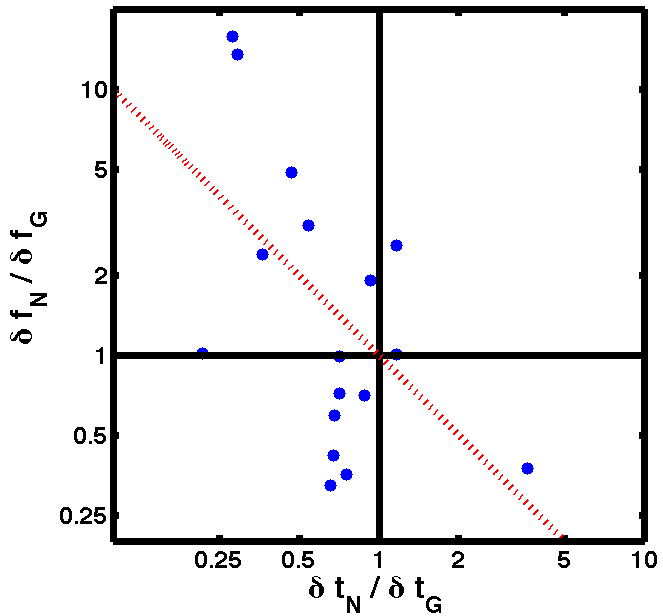}
\end{center}
\caption{{\bf Difference in performance on different shaped notes.}  On the left is performance on the ``time-reversed" pulse divided by performance on the ``time-forward" notelike pulse.  On the right is performance on the ``notelike" pulse divided by performance on the gaussian pulse.  The dotted red line indicates a perfect tradeoff in performance, i.e. $\frac{\delta t_1}{\delta t_2} \frac{\delta f_1}{\delta f_2} = 1$.  On both figures, axes are logarithmic.  \label{VectorDifPlot}
}\end{figure}

In Figure \ref{TF Plot}, we plot our data on a time-frequency plane with logarithmic axes: red dots indicate results from notelike pulses; green x's, time-reversed.  Note that the data from the time-reversed pulses seem to cluster on the right-hand side of the plot.  To quantify this effect, we examined the vector of change in performance between the notelike and the time-reversed notelike pulses (Figure \ref{VectorDifPlot}).   Note the dramatically worse performance in timing, frequency, or both.  A compensatory improvement in timing acuity occurred only twice and of relatively small multiplicative effect, whereas an improvement  in frequency discrimination was seen four times.  Performing a t-test on the fold-change in performance in frequency and timing shows there is no statistically significant effect in frequency ($P > 0.76$), but a strongly significant one in timing ($P < 0.0014$). 

As a methodological control, we combined our data with that of our previous study \cite{OppenheimMagnasco} and examined the difference in simultaneous time-frequency acuity between gaussian and notelike pulses.  Out of 15 pairs of tests, 12 subjects showed improved performance in timing and 8 in frequency.  Applying the same t-test to the difference in log-acuity, we found a significant change in $\delta t$ ($P < 0.018$), but no significant change in $\delta f$ ($P >  0.29$) or in the uncertainty product ($P > 0.66$).  

\section{Bounds on Nonlinearity}

The results of this control confirm our prior conclusion: that the Fourier Uncertainty Principle does not limit human-time frequency perception \cite{OppenheimMagnasco}.  Nonetheless, many nonlinear time-frequency distributions obey a modified uncertainty relation, changing only the prefactors of $\Delta t$ and $\Delta f$ \cite{CohenBook}.  Given that a $\sim 5.7$-fold increase in the theoretical uncertainty product did not change the measured product, $\delta t \delta f$ and that the best subject measured beat the uncertainty principle by a factor of $\sim 50$, we may place a bound on the magnitude of the prefactors in any ``modified" uncertainty relation and thus the order of the nonlinearity of any distribution used in auditory processing. As there is no reason to believe a general closed form exists for arbitrary pulses, we focus below on Gaussian wavepackets.  For mathematical simplicity, we work with the angular frequency, $\omega = 2\pi f$.

We may define a hierarchy of nonlinear time-frequency distributions based on the order of their nonlinearity: the number of the copies of the signal that are convolved in the transformation.  All time-frequency representations may be written in terms of Cohen's class \cite{CohenBook}:

\begin{eqnarray*}
C(t,\omega) &=& \frac{1}{4\pi^2} \int\int\int e^{-i\theta t -i\tau\omega+i\theta u}\Phi(\theta,\tau) \\
& & \times s^*(u-\frac{1}{2}\tau) s(u+\frac{1}{2}\tau)\quad \mathrm{d}u \mathrm{d}\tau \mathrm{d}\theta
\end{eqnarray*}
Here, $t$ and $\omega$ are our time and frequency variables in the final representation, $C$; $s$ is the original signal, $\tau$ the running window in time, centered at $u$, and $\theta$ the frequency window.  $\Phi$ is denoted the kernel of the transformation.  For all bilinear time-frequency representations (Cohen's Class), $\Phi$ is independent of $s$.  The simplest member of this class is the Wigner Distribution, for which $\Phi = 1$.  A signal-dependent kernel changes the order of the distribution\footnote{One may recover the spectrogram by inserting a window-dependent kernel and writing $C(t,\omega)$ as a square}.   To obtain properly-construed marginal distributions in time and frequency and thus values for $\Delta t$ and $\Delta f$, 

\begin{eqnarray*}
\int C(t,\omega)\mathrm{d}t &=& |S^*(\omega)|^2 \\
\int C(t,\omega)\mathrm{d}\omega &=& |S(t)|^2
\end{eqnarray*}
we must have $\Phi(0,\tau) = 1$ and $\phi(\theta,0) = 1$.  Hence, any time-frequency representation whose kernel obeys the above conditions must obey a modified uncertainty principle of the form $K\Delta t \Delta f \geq \frac{1}{4\pi}$, with K some constant, dependent on both the wavepacket shape and the enveloped frequencies.  

To measure the increase in precision, we evaluate the Cohen's class integrals for gaussian-enveloped pure tones, $s = e^{-t^2/4\sigma^2}e^{i\omega t}$, both because of their optimality in the linear case and our data on such packets.  For $\Phi = 1$ (the Wigner Distribution), we have after taking the Fourier transform of $\theta$ and integrating out $u$,
\begin{equation}
C(t,\omega) = \int e^{-i\tau\omega} e^{-t^2/2\sigma^2 -\tau^2/8\sigma^2} \mathrm{d}\tau
\end{equation}
Without evaluating this integral, we may read off $\Delta t = \sigma$ and $\Delta f = \frac{1}{2\pi}\Delta \omega = \frac{1}{2\pi}\frac{1}{4\sigma}$, a factor of two improvement over linearity.  Examining the general case more carefully, we see that the origin of improved performance is due to the convolution of the signal with itself.  Without any prior information, $\Phi$ is not a function of $u$, $t$, or $\omega$.  The kernel is only able to affect the uncertainty in the frequency domain by action on $\theta$.  In the case of Gaussian wavepackets, where enhanced precision arises from the largeness of N in the Fourier transform of $e^{-\tau^2/N\sigma^2}$, the only kernel that could improve upon the Wigner distribution would be of the form $e^{\tau^2/M\sigma^2}$ with $M<N$.  Such a kernel, by privileging large $\tau$'s over small ones would be catastrophic in the case of multiple signal components, amplifying the destructive interference of the Wigner Distribtuion.  The kernels used in time-frequency analysis tend to be sharply peaked to decrease this interference.  A paradigmatic example is the Choi-Williams Kernel, $e^{-a\theta^2\tau^2}$ which applies a Gaussian window in both time and frequency to the signal autocorrelation, thus reducing the enhanced precision in the frequency domain from convolving the signal with itself.  

We may apply the same logic to kernels that are explicitly dependent on the signal itself.  Each additional pair of the signal, $ s^*(u-\frac{1}{2}\tau) s(u+\frac{1}{2}\tau)$, will build higher order correlations into $C(t,\omega)$, increasing precision by a factor of 2.  Any other terms will be used to suppress interference (increasing resolution) and cannot add to precision.  Our best subject was able to beat the Uncertainty Principle by a factor of $~50$, suggesting at a minimum, a 12th order nonlinearity (6 factors of 2, each arising from a signal-conjugate pair) is necessary to explain human auditory perception.  Such a bound rules out the traditional Cohen's Class representations and throws into doubt the Hilbert-Huang method and other PCA-based methods, which, involving a matrix-inversion, can be estimated to be of greater than 3rd but less than 12th order \cite{HilbertHuang}.  Our results thus suggest that only matching-pursuit, the multi-tapered spectral derivatives \cite{Thomson, Mitra} and the reassigned spectrograms \cite{KoderaVilledary, AugerFlandrin, Flandrin:DR, Fitz, GardnerMagnasco} can hope to capture the precision of the auditory system.  

\section{Time-Reversal Symmetry Breaking}
The significant increase in timing acuity unaccompanied by a drop in the uncertainty product found for a pulse of considerably larger theoretical uncertainty in both timing and frequency indicates that either the precision of human time-frequency perception operates in a realm distant from the true uncertainty bound, or such a bound does not exist for the auditory system.  We may increase both the physical $\Delta t$ and $\Delta f$ of a note, shaping the envelope to aid in temporal perception, and find improved timing acuity without a decrease in simultaneous time-frequency acuity.  Examining Figure \ref{VectorDifPlot}, in the right panel, we see data both above and below the line $\delta f = 1/\delta t$, which represents a perfect tradeoff between time and frequency.  In the right panel, we see that the best performers fall almost exactly on this line, indicating a perfect tradeoff between time and frequency acuity when going from ``notelike" to time-reversed pulses; every other subject is above this line.  These are the subjects whose statistically significant worse performance indicates time-reversal symmetry breaking in auditory processing.

The observed asymmetry in auditory processing takes our experimental and theoretical results on the uncertainty principle one step further.  The indifference of the Fourier Transform to the arrow of time is not reproduced in perceptual acuity.  A weaker version of time-reversal symmetry might suggest that overall time-frequency acuity for notes is the same in both the time-forward and time-reversed cases, with increased temporal acuity in the ``time-forwards" direction trading off with increased frequency acuity in the ``time-reversed" direction.  While some subjects are able to manage this feat to a certain extent, it is by no means uniform across subjects.  Compensatory improvements in frequency perception may not explain all of their performance; in many subjects we have noticed an increase in performance on the simultaneous time-frequency discrimination task with exposure for the first 2-3 trials.  It is likely this is the origin of the improved timing perception of the one subject who could distinguish notes in time better in the time-reversed case than the original one.  To properly disentangle which effects are due to repeated exposure and which are due to innate statistical priors in the auditory system---those which are built from years of exposure to time-reversal symmetry broken sounds and  those which are due to the constraints of auditory physiology---requires further study and careful controls for task learning.

We have demonstrated that human auditory perception is primed for the shapes of natural sounds, a sharp attack followed by a long decay, corresponding to the physics of natural sound production.  We have used simple, direct psychophysical measurements to test for the changes in simultaneous time-frequency acuity after reversing the temporal direction of symmetry-broken pulses, lending credence to, at the minimum, statistical priors for sharp attack, long decay sounds.  Such statistical priors add to the growing body of evidence that human auditory processing is adapted for natural sounds. Not only then is auditory processing inherently nonlinear, these nonlinearities are used to improve perceptual acuity to sounds that correspond to the physics of natural sound production. 

Our experimental results inspired a look at the hierarchy of nonlinear time-frequency distributions and allowed the placement of a lower bound on the degree of such a nonlinearity, ruling out many of the simplest and most frequently used nonlinear time-frequency representations, the bilinear ones of Cohen's Class, as well as suggesting that ones based on PCA, such as Hilbert-Huang are not of high enough order to account for our data.  Among those that remain, the spectral derivatives and the reassigned spectrograms, we hope that further psychophysical measurements as well as considerations of the ability to implement such algorithms in neural ``hardware" may further narrow the class of plausible methods of auditory processing.  Lastly, our observations about time-reversal symmetry breaking and the temporal precision of the auditory system suggest further research into this ecologically-relevant domain.

\section*{Acknowledgments}
Supported in part by NSF grant EF-0928723 and by the Simons Foundation.  We gratefully acknowledge the musical expertise of Deniz Hughes throughout this effort.

\bibliography{TRbib}

\section*{Figure Legends}

\end{document}